\documentclass[a4paper,11pt]{article}
\pdfoutput=1 
\usepackage{jcappub} 
\usepackage{aasmacros}              
\usepackage{comment}
\usepackage{lscape}
\usepackage{verbatim}
\usepackage{comment}
\usepackage{array}
\usepackage{subcaption}
\usepackage{multirow}
\usepackage{graphicx}
\graphicspath{{graphics/}}
\usepackage{amsmath}
\usepackage{amssymb}
\usepackage{rotating} 
\usepackage{xspace}
\usepackage[dvipsnames]{xcolor}
\usepackage{booktabs}
\usepackage[normalem]{ulem}
\usepackage{adjustbox}
\usepackage{lipsum}  
\usepackage{caption}
\usepackage{placeins}
\usepackage{afterpage}

\newcommand{\Msun}{{\rm M}_\odot}
\renewcommand{\epsilon}{\varepsilon}

\title{LMC-Perturbed LZ Dark Matter Landscape}

\author[a,b]{Nassim Bozorgnia,}
\author[c]{Samantha Contreras,}
\author[c]{Graciela B. Gelmini,}
\author[c]{Alvine C. Kamaha,}
\author[d]{Javier Reynoso-Cordova,}
\author[c,e]{and Yongheng Xu}

\affiliation[a]{Department of Physics, University of Alberta,
CCIS 4-181, \\ Edmonton, Alberta T6G 2E1, Canada}
\affiliation[b]{Theoretical Physics Institute, University of Alberta, CCIS 4-181,\\ Edmonton, Alberta, Canada}
\affiliation[c]{Department of Physics and Astronomy, UCLA, \\
475 Portola Plaza, Los Angeles, CA 90095, USA}
\affiliation[d]{Istituto Nazionale di Fisica Nucleare, Sezione di Napoli, 
\\Complesso Universitario di Monte Sant'Angelo,
Via Cintia, 80126 Napoli, Italy}
\affiliation[e]{Fysisk institutt, University of Oslo, Sem S{\ae}landsvei 24, N-0371, Oslo, Norway}

\emailAdd{nbozorgnia@ualberta.ca}
\emailAdd{scont@g.ucla.edu}
\emailAdd{gelmini@physics.ucla.edu}
\emailAdd{akamaha@physics.ucla.edu}
\emailAdd{javier.reynoso@na.infn.it}
\emailAdd{xuyongheng@physics.ucla.edu}

\abstract{
The local dark matter (DM) velocity distribution is significantly altered by the gravitational impact  of the Large Magellanic Cloud (LMC), which creates a high-velocity tail. We evaluate how the LMC  reshapes the DM landscape for a wide variety of theoretical models in light of the recent  putative LZ DM event. We show that the LMC-induced velocity shifts substantially modify the parameter space for inelastic endothermic DM models across a broad range of possible interactions, driving the viable regions towards larger mass splittings, which are more susceptible to collider and indirect detection constraints. By contrast, models for elastically colliding DM  are less impacted, and some remain viable candidates -- for example, the simple light Singlet-Doublet Majorana DM candidate at the Higgs blind spot. Our work confirms that accurately accounting for LMC effects is essential when high-velocity DM is probed, rather than relying on the simplified Standard Halo Model.
}

\begin{document}
\maketitle
\flushbottom

\section{Introduction}
\label{sec:intro}
Identifying dark matter (DM) is one of the most critical objectives in particle physics and cosmology today~\cite{cirelli2026dark, Bozorgnia:2024pwk}. Direct detection seeks to observe the energy deposited in highly sensitive underground-based detectors by  DM particles~\cite{Goodman:1984dc,Ahlen:1987mn,Lewin:1995rx,PhysRevD.33.3495}.

The LUX-ZEPLIN (LZ) direct detection experiment recently reported an unusual event, LZ230616, with a reconstructed recoil energy of $248\pm23_{\rm stat}\pm23_{\rm sys}$ keV~\cite{LZ:2026axp}. The event was found in a 2.84 tonne-year search that extended the nuclear-recoil (NR) energy window from 5.4 keV to 270 keV, beyond that used in conventional WIMP analyses. LZ searches for DM and other rare
signals~\cite{LZ:2023FirstDM,LZ:2024zvo,
LZ:2024UltraheavyDM,LZ:2025CosmicRayBoostedDM,
LZ:2023LowEnergyER,LZ:2026LowEnergyER,
LZ:2026SolarNeutrinos,LZ:2025AtmosphericMCP,LZ:2025DoubleElectronCapture} with a dual-phase xenon time projection chamber (TPC) located a mile underground at the Sanford Underground Research Facility. Combined measurement of scintillation and ionization signals in the TPC allows the reconstruction of interaction energies and positions, as well as discrimination between NRs and electron-recoil backgrounds. The isolated event passed the analysis selections and is consistent with a single nuclear recoil within the fiducial volume, in a region where the expected background is very low. Dedicated studies of rare backgrounds and detector effects did not identify a likely explanation. Although this observation is insufficient to establish a DM origin, its measured properties can be accommodated by certain DM interaction models, motivating further investigation of this possibility and tests with additional data.

The single  LZ high-energy nuclear recoil event has prompted a flurry of DM interpretations. A panorama of different ideas, necessarily with an incomplete set of references, includes the following.  

Inelastic endothermic DM, in which the initial DM particle of mass $m_\chi$ scatters into another with a larger mass, $m_\chi + \delta$, is the most widely adopted interpretation because the kinematic threshold in these models suppresses low-energy recoils, naturally fitting the isolated high-energy event without producing a low-energy tail. Example realizations include electroweak-doublet higgsino-like DM of about 1 TeV of mass~\cite{Freese:2026sga,Wu:2026nhi,DiMauro:2026ldr,Gemmell:2026yaw} and five-dimensional models where the splitting has a geometric origin~\cite{Ahmed:2026qjg,Ahmed:2026com}. Inelastic DM pushed to operate close to the Galactic escape speed could produce a seasonal strong signal due to a large annual modulation~\cite{McCabe:2026crm}.  Several global kinematic recasts of the inelastic parameter space show that, because of their dependence on high speed DM, they are susceptible to the effects of the Large Magellanic Cloud (LMC)~\cite{Fan:2026kxx,Wu:2026nhi,Ghosh:2026txe,OHare:2026nqi, Rodd:2026tyn, He:2026hqz, Asadi:2026iot}.  The inelastic exothermic interpretation is tightly constrained by collider limits~\cite{Du:2026lpa,Okada:2026fef,Maier:2026qpr},  as well as indirect detection via continuous gamma-ray searches by Fermi-LAT and HESS~\cite{Wu:2026nhi}, and IceCube and Super-Kamiokande high energy neutrino searches~\cite{Ghosh:2026txe, Bose:2026szs, Pospelov:2026scg}. 

Inelastic exothermic~\cite{Dent:2026bji,deLima:2026shq,Baer:2026fpy} and asymmetric~\cite{Nagata:2026pbj} DM have a metastable DM state which down-scatters into a lighter state releasing a fixed amount of stored energy into the nuclear recoil. These models produce a sharp, localized peak in the spectrum that could explain the LZ event. They escape indirect detection limits, but are challenged by  missing energy constraints at colliders and cosmological lifetime thresholds.

Also, Composite~\cite{Sheng:2026tqt,Jung:2026otm} and Dark QCD~\cite{Sannino:2026hkc} DM scenarios with multi-component composite states or dark hadrons, can produce high-energy recoils through inelastic  transitions between excited dark-hadronic states, dark baryonic number violation, or the exchange of light dark-sector portals (like an axion portal~\cite{An:2026pkc}) that modify the nuclear form factors at high momentum transfer. While these composite models successfully circumvent standard electroweak direct-annihilation bounds, they are tightly constrained by LHC dijet and emerging jet searches for dark hadrons, as well as astrophysical limits on the accompanying axion and light gauge portals.

Elastic collisions of DM with spin-independent (SI) predict too many events at low energy to be viable, but other non-relativist effective field theory (NREFT) operators are compatible with the LZ event. Operators that lead to momentum suppressed or spin-dependent (SD) inherently shift the expected nuclear recoil spectrum toward higher energies and remain viable~\cite{Palmisano:2026kuj,Arcadi:2026kev}. Also boosted DM with SD interactions has been explored~\cite{Alhazmi:2026efz,Chauhan:2026udz}. These models are challenged by Solar capture and Neutrino Telescope limits~\cite{Bose:2026szs}.

In this paper we investigate how the contribution of the LMC to the local DM velocity distribution reshapes the DM landscape to explain the high recoil energy and momentum of the LZ event, for a variety of NREFT interaction operators and elastic and inelastic scattering.  We study the resulting DM parameter space for some preferred models. We also investigate the viability of an elastically scattering,  relatively light Majorana Singlet-Doublet DM  candidate, at the Higgs blind spot as a simple yet viable model.

The importance of the LMC in this context follows from the significant role of the local DM phase space distribution in interpreting direct detection results.
A commonly used benchmark is the Standard Halo Model (SHM)~\cite{PhysRevD.33.3495}, which describes the Milky Way (MW) halo as an isotropic, isothermal sphere with a Maxwell-Boltzmann velocity distribution in the Galactic frame. High-resolution cosmological simulations show that the local velocity distribution of the smooth DM component in MW-like halos is generally well described by a Maxwellian. At the same time, significant halo-to-halo variations are found across simulated galaxies, introducing substantial astrophysical uncertainties in the local DM phase space distribution and the resulting interpretation of experimental data~\cite{Bozorgnia:2016ogo, Kelso:2016qqj, Sloane:2016kyi, Bozorgnia:2017brl, Bozorgnia:2019mjk, Poole-McKenzie:2020dbo, Lawrence:2022niq, Kuhlen:2013tra, Lacroix:2020lhn, Santos-Santos:2023ubx}. Cosmological hydrodynamical simulations of the Local Group also find no evidence for a distinct population of high-speed extragalactic DM particles in the Solar neighborhood~\cite{Santos-Santos:2023ubx}.

A key departure from the SHM is the modification of the local DM phase space induced by the recent infall of the LMC. As the most massive satellite of the Milky Way, the LMC generates substantial perturbations in both the stellar and DM components of the Galaxy~\cite{GaravitoCamargo:2021tcp, Garavito-Camargo:2020lqm, Conroy_2021, Petersen_2020, Peterson_2020_MNRAS, Cunningham:2020nlo, Cavieres_2025}. Studies based on both idealized~\cite{Besla:2019xbx,Donaldson:2021byu} and cosmological~\cite{Smith-Orlik:2023kyl} simulations find that these perturbations can substantially boost  the high speed tail of the DM velocity distribution in the Solar neighborhood. The resulting enhancement can have important consequences for direct detection, shifting exclusion limits by several orders of magnitude toward lower cross sections and DM masses~\cite{Besla:2019xbx, Smith-Orlik:2023kyl}. Such effects are particularly relevant for velocity-dependent interactions and inelastic DM~\cite{Reynoso-Cordova:2024lmc}. The LMC also extends the reach of direct detection searches for ultraheavy DM well above the TeV scale~\cite{Bozorgnia:2025lsl}. Additionally, its perturbation of the local DM velocity distribution can significantly modify the predicted directional signatures of DM~\cite{Reynoso-Cordova:2026pri}.

The outline of the paper is as follows. In section~\ref{sec:lmc}, we describe the simulated MW+LMC analogue used in this work and discuss its impact on the local DM velocity distribution. In section~\ref{sec:rates}, we describe the calculation of direct detection event rates, while section~\ref{sec:eft} introduces the NREFT interaction operators. In section~\ref{sec:stat}, we discuss the details of our statistical analysis. In section~\ref{sec:region} we present the best fit parameter space for the NREFT interactions. In section~\ref{sec:SDM}, we describe the Singlet-Doublet model and its corresponding favored parameter space. Finally, in section~\ref{sec:summary}, we summarize our conclusions.

\section{LMC perturbed local dark matter halo}
\label{sec:lmc}

In this work, we use the data of the simulated MW+LMC system originally identified  in ref.~\cite{Smith-Orlik:2023kyl}. The system is obtained from the Auriga magneto-hydrodynamical simulations~\citep{Grand:2016mgo, Grand:2024}, which are a set of zoom-in simulations of isolated MW mass halos selected from a periodic volume of ($100$~Mpc)$^3$ (L100N1504) from the EAGLE project~\cite{Schaye:2014tpa, Crain:2015poa}.  The simulations were performed using the moving-mesh code Arepo~\citep{Springel:2009aa} and incorporate a galaxy formation subgrid model which includes black hole formation, star formation, active galactic nuclei and supernova feedback, metal cooling, and background UV/X-ray photoionisation radiation~\cite{Grand:2016mgo}. The  simulations reproduce a range of observed properties of present day MW–mass systems, including their stellar masses and sizes, rotation curves, star formation rates, and metallicities. The simulations adopt the Planck-2015~\citep{Planck:2015fie} cosmological parameters: $\Omega_{m}=0.307$, $\Omega_{\rm bar}=0.048$, $H_0=67.77~{\rm km~s^{-1}~Mpc^{-1}}$. Our analysis uses the standard resolution level (Level 4) of the Auriga simulations, with a DM particle mass of $m_{\rm DM} \sim 3\times 10^5~\Msun$, a baryonic particle mass of $m_b=5\times10^4~\Msun$, and a Plummer equivalent gravitational softening length of $\epsilon=370$~pc~\citep{Power:2002sw,Jenkins2013}.

We adopt the MW+LMC  analogue based on the re-simulated halo 13 of ref.~\cite{Smith-Orlik:2023kyl}, corresponding to the Auriga 25 halo and its associated LMC analogue. This halo was re-simulated with increased snapshots around the LMC analogue's pericentric passage. The MW analogue has a virial mass\footnote{The virial mass is defined as the mass enclosed within the radius at which the mean enclosed matter density is 200 times the critical density of the Universe.} of $1.2 \times 10^{12}~\Msun$, while the LMC analogue has a virial mass of $3.2 \times 10^{11}~\Msun$ at infall. A detailed description of the system and its selection is given in ref.~\cite{Smith-Orlik:2023kyl}. In our analysis, we use the snapshot closest to the present-day MW--LMC separation. At this snapshot, the LMC analogue is located approximately $50$~kpc from the MW analogue and has a relative speed of 317~km/s, in close agreement with the observed MW--LMC system~\cite{Besla:2007kf, Salem2015, Kallivayalil:2013xb}. 

The position and velocity of the  Sun within the simulated halo are chosen to reproduce the observed Sun-LMC geometry, following the procedure described  in ref.~\cite{Smith-Orlik:2023kyl}. We define the \emph{Solar region} as the intersection between a spherical shell extending from 6 to 10~kpc from the center of the MW analogue, encompassing the Solar radius of $\sim 8$~kpc, and a cone with an opening angle of $\pi/4$ radians. The cone is centered on the chosen Solar position and has its vertex at the galactic center (see figure~2 of ref.~\cite{Smith-Orlik:2023kyl}). This choice provides a sufficiently localized region to be sensitive to the chosen Solar position, while still containing several thousand DM particles, as discussed in ref.~\cite{Smith-Orlik:2023kyl}.

\begin{figure}[t]
    \centering
        \includegraphics[width=9cm]{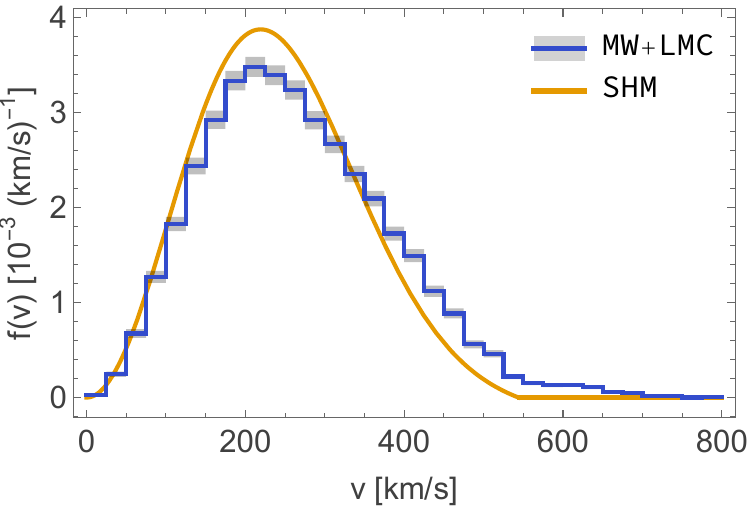} 
    \caption{The local DM speed distribution in the Galactic rest frame for the simulated MW+LMC analogue (blue) and the SHM (orange). The shaded gray band corresponds to the 1$\sigma$ Poisson errors in the speed distribution from the simulated system.}
\label{fig:fv}
\end{figure}

The percentage of DM particles originating from the LMC in the Solar region is 0.26~\cite{Smith-Orlik:2023kyl}. This percentage is defined as the ratio of the number of DM particles originating from the LMC and the total number of DM particles in the Solar region, multiplied by 100. Despite this small percentage, the LMC significantly shifts the high-speed tail of the local DM velocity distribution to higher speeds. 

Figure~\ref{fig:fv} shows a comparison of the local DM speed distribution in the Galactic rest frame for the simulated MW+LMC analogue and the SHM. The speed distributions are defined as $f(v) = v^2 \int d\Omega_{{\bf v}}\Tilde{f}({\bf v})$, where $\Tilde{f}({\bf v})$ is the local DM velocity distribution normalized to 1, such that $\int dv f(v)=\int d^3v \Tilde{f}({\bf v}) = 1$, and $d\Omega_{{\bf v}}$ is an infinitesimal solid angle around the direction ${\bf v}$. The orange curve shows a Maxwellian speed distribution with peak speed of 220~km/s, truncated at a Galactic escape speed of 544~km/s, as commonly adopted in the SHM. The blue curve shows the local DM speed distribution of the simulated MW+LMC analogue, with the gray shaded band corresponding to 1$\sigma$ Poisson errors in the speed distribution. To compare directly with the SHM, the DM speeds extracted from the simulations are scaled by $(220~{\rm km/s})/v_c$, where $v_c$ is the local circular speed of the MW analogue, calculated from the total mass within a sphere of radius 8~kpc. 

We can clearly see the impact of the LMC on the high-speed tail of the local DM speed distribution in figure~\ref{fig:fv}. While the speed distribution vanishes above the Galactic escape speed in the SHM, it extends to speeds of $\sim 800$~km/s for the MW+LMC. The high-speed tail is particularly important for endothermic inelastic DM scattering, since the DM mass splitting raises the minimum speed required to produce a nuclear recoil in a direct detection experiment, so that only the fastest DM particles can scatter (see eq.~\eqref{eq:vmin_inel} below).

To compute direct detection event rates, we need to transform the local DM velocity distribution from the Galactic reference frame to the  detector's reference frame,  $\tilde{f}_{\rm{det}}({\bf v},t)=\tilde{f}_{\rm{gal}}({\bf v} + {\bf v}_s + {\bf v}_e(t))$, where ${\bf v}_s={\bf v}_c + {\bf v}_{\rm{pec}}$ is the velocity of the Sun with respect to the Galactic center, ${\bf v}_c$ is the Sun's circular velocity, for which we adopt ${\bf v}_c = (0, 220, 0)$~km/s, ${\bf v}_{\rm{pec}}=(11.10,12.24,7.25)$~km/s~\cite{2010MNRAS.403.1829S} is the Sun's peculiar velocity in Galactic coordinates, and ${\bf v}_e(t)$ is the Earth's velocity with respect to the Sun. For simplicity, we neglect the small eccentricity of the Earth's orbit.

\section{Direct detection event rates}
\label{sec:rates}

We consider a DM particle of mass $m_\chi$ scattering with a target nucleus of mass $m_T$ in an underground detector, and depositing the nuclear recoil energy, $E_R$.  If the detector includes multiple  nuclides, the differential event rate per  unit energy, detector mass, and time is given by 
\begin{equation}
    \frac{dR}{dE_R}=\sum_T \frac{dR_T}{dE_R} =\sum_T \frac{C_T}{m_T}\,\frac{\rho_\chi}{m_\chi}\int_{v\geq v_{{\rm min}}} d^3 v\, \frac{d\sigma_T}{dE_R}\,v\, \Tilde{f}_{\rm det}({\bf v}, t)~,
    \label{Eq: dRdER}
\end{equation}
where the sum is over the different isotopes or target nuclides, $T$, in the detector, $C_T$ is the mass fraction of $T$ in the detector, $d\sigma_T/dE_R$ is the differential DM-nucleus scattering cross section for nuclide $T$, $\rho_\chi$ is the local DM density, ${\bf v}$ is the relative velocity between the DM and the target nucleus, with $v\equiv |{\bf v}|$, and $\Tilde{f}_{\rm det}({\bf v}, t)$ is the local DM velocity distribution in the Earth's rest frame.  

For elastic scattering, the minimum speed required for the DM particle to produce a recoil energy $E_R$ in the detector is
\begin{equation}
    v_{{\rm min}}=\sqrt{\frac{m_TE_R}{2\mu_{\chi T}^2}}~,
\end{equation}
where $\mu_{\chi T}$ is the DM-nucleus reduced mass.

For inelastic DM scattering~\cite{Tucker-Smith:2001myb}, the DM particle $\chi$ scatters to an excited state $\chi^\ast$, with a mass splitting $\delta=m_{\chi^\ast}-m_\chi$. We consider the endothermic case, for which $\delta>0$. We have
\begin{equation}
 v_{{\rm min}}=\sqrt{\frac{1}{2m_T E_R}}\left(\frac{m_T E_R}{\mu_{\chi T}}+\delta\right).
\label{eq:vmin_inel}
\end{equation}
The dependence of $v_{\rm{min}}$ on the mass splitting, $\delta$, increases the minimum DM speed required to produce a recoil signal.

For the standard SI and SD DM-nucleus interactions, the differential cross section is proportional to $v^{-2}$ and the  event rate becomes proportional to the {\it halo integral},
\begin{equation}
    \eta(v_{{\rm{min}}},t) \equiv \int_{v > v_{{\rm{min}}}} d^3v \, \frac{\Tilde{f}_{\rm{det}}({{\bf v}},t)}{v},
\label{eq:etavmin}
\end{equation}
which together with the local DM density, $\rho_\chi$, encompass the astrophysical dependence of the event rate. 

For a general set of non-relativistic effective interactions discussed in section~\ref{sec:eft}, the differential cross section can be expressed as a linear combination of a velocity-dependent term proportional to $v^{-2}$ and a velocity-independent term~\cite{Barger:2010gv, DelNobile:2012tx, DelNobile:2015tza}. The former leads to the halo integral, $\eta(v_{\rm min}, t)$ (eq.~\eqref{eq:etavmin}), while the later results in another velocity integral defined as 
\begin{equation}
    h(v_{{\rm{min}}},t) \equiv \int_{v>v_{{\rm{min}}}} d^3 v\, v\, \tilde{f}_{\rm{det}}({{\bf v}},t).
\label{eq:hvmin}
\end{equation}
The impact of the LMC on the local DM velocity distribution will affect both $\eta(v_{{\rm min}}, t)$ and $h(v_{{\rm min}}, t)$. 

\section{Non-relativistic effective interactions}
\label{sec:eft}
The NREFT framework provides a systematic description of the possible contact interactions between DM and nucleons that can arise from an underlying theory of DM~\cite{Fan:2010gt, Fitzpatrick:2012ix, Fitzpatrick:2012ib, Anand:2013yka, Dent:2015zpa}. It generalizes the conventional SI and SD interactions by allowing for operators with explicit dependence on the momentum transferred from DM to the nucleon and the DM-nucleon relative velocity. The DM-nucleus interaction can  be expressed in terms of a basis of independent operators, $\mathcal{O}_i$. We restrict our analysis to one-body DM-nucleon interactions mediated by a heavy spin-0 or spin-1 particle, and assume that the DM particle has spin $1/2$.
The operators $\mathcal{O}_i$ as given in ref.~\cite{Fitzpatrick:2012ix} form a generalized non-relativistic interaction Lagrangian. These  are
\begin{equation}
\begin{aligned}
&\mathcal{O}_1 = \mathbf{1}_\chi \mathbf{1}_N, \quad
\mathcal{O}_2 = (\mathbf{v}^\perp)^2, \quad
\mathcal{O}_3 = i \mathbf{S}_N \cdot \left( \frac{\mathbf{q}}{m_N} \times \mathbf{v}^\perp \right), \\
&\mathcal{O}_4 = \mathbf{S}_\chi \cdot \mathbf{S}_N, \quad
\mathcal{O}_5 = i \mathbf{S}_\chi \cdot \left( \frac{\mathbf{q}}{m_N} \times \mathbf{v}^\perp \right), \quad
\mathcal{O}_6 = \left( \mathbf{S}_\chi \cdot \frac{\mathbf{q}}{m_N} \right) \left( \mathbf{S}_N \cdot \frac{\mathbf{q}}{m_N} \right), \\
&\mathcal{O}_7 = \mathbf{S}_N \cdot \mathbf{v}^\perp, \quad
\mathcal{O}_8 = \mathbf{S}_\chi \cdot \mathbf{v}^\perp, \quad
\mathcal{O}_9 = i \mathbf{S}_\chi \cdot \left( \mathbf{S}_N \times \frac{\mathbf{q}}{m_N} \right), \\
&\mathcal{O}_{10} = i \mathbf{S}_N \cdot \frac{\mathbf{q}}{m_N}, \quad
\mathcal{O}_{11} = i \mathbf{S}_\chi \cdot \frac{\mathbf{q}}{m_N}, \\
&\mathcal{O}_{12} = \mathbf{S}_\chi \cdot \left( \mathbf{S}_N \times \mathbf{v}^\perp \right), \quad
\mathcal{O}_{13} = i \left( \mathbf{S}_\chi \cdot \mathbf{v}^\perp \right) \left( \mathbf{S}_N \cdot \frac{\mathbf{q}}{m_N} \right), \\
&\mathcal{O}_{14} = i \left( \mathbf{S}_\chi \cdot \frac{\mathbf{q}}{m_N} \right) \left( \mathbf{S}_N \cdot \mathbf{v}^\perp \right), \quad
\mathcal{O}_{15} = - \left( \mathbf{S}_\chi \cdot \frac{\mathbf{q}}{m_N} \right) \left[ \left( \mathbf{S}_N \times \mathbf{v}^\perp \right) \cdot \frac{\mathbf{q}}{m_N} \right].
\end{aligned}
\label{eq:NR_operators}
\end{equation}
Here ${\bf q}$ is the momentum transferred from the DM to the nucleon, ${\bf v}^\perp$ is  the transverse   DM-nucleon relative velocity -- defined as the component of the relative velocity orthogonal to the momentum transfer-- ${\bf v}^\perp \cdot {\bf q} = 0$, ${\bf S}_\chi$ and ${\bf S}_N$ are the DM and nucleon spin operators, respectively, and $m_N$ is the nucleon mass.

The operators $\mathcal{O}_1$ and $\mathcal{O}_4$ correspond to the SI and SD couplings, respectively. We neglect operator $\mathcal{O}_2$ in this analysis, as it is quadratic in ${\bf v}^\perp$ and does not arise at leading order in the nonrelativistic limit of a relativistic interaction~\cite{Fitzpatrick:2012ix}.

For many of these operators, the DM-nucleon cross section has a different dependence on the relative DM velocity, compared to $\mathcal{O}_1$ and $\mathcal{O}_4$. As a result, modifications to the local DM velocity distribution induced by the LMC can have a larger effect on direct detection constraints for the corresponding effective couplings~\cite{Reynoso-Cordova:2024lmc}. In the case of endothermic inelastic DM scattering, direct detection experiments are  particularly sensitive to the high-speed tail of the local velocity distribution. The impact of the LMC on direct detection constraints can therefore be especially significant in this scenario~\cite{Reynoso-Cordova:2024lmc}.

The effective DM--proton and DM--neutron couplings for each operator $\mathcal{O}_i$ are denoted as $c_i^p$ and $c_i^n$, respectively. In the isospin representation~\cite{Fitzpatrick:2012ix}, we have\footnote{Notice that we follow the normalization used in ref.~\cite{Fitzpatrick:2012ix} and in \texttt{DDCalc}, which differs from the one used in ref.~\cite{Anand:2013yka} where $c_i^s =(c_i^p + c_i^n)/2$.}
\begin{align}
c_i^s &=c_i^p + c_i^n\,, \hspace{5em}  c_i^v =c_i^p - c_i^n\,,
\end{align}
where $c_i^s$ and $c_i^v$ are the isoscalar and isovector couplings to each operator, respectively, and have dimensions of $1/{\rm (mass)}^2$. In section~\ref{sec:region}, we consider purely isoscalar or purely isovector
interactions, setting $c_i^v = 0$ or $c_i^s = 0$, respectively. 

\section{Statistical analysis}
\label{sec:stat}
In xenon TPCs like LZ, the energy of an event, which we call $E'$, is reconstructed from the sizes of the scintillation signal (S1) and ionization signal (S2).
A nuclear recoil of energy $E_R$ produces on average a number of scintillation photons and ionization electrons fixed by the NR yield model. The S1 and S2 signals fluctuate about these means because of fluctuations in quenching, quanta production and signal detection. Whether an event passes the trigger, the S1 threshold and the analysis selections, and how far its reconstructed energy falls from $E_R$, are both outcomes of these fluctuations. These two effects are therefore described by two functions of $E_R$: the efficiency $\epsilon(E_R)$, defined as the probability that a recoil of energy $E_R$ is selected, and the energy resolution function $G_T(E', E_R)$, defined as the probability density for a selected recoil if energy $E_R$ to be reconstructed as having energy $E'$. For the event, we take $E_R=$ 248 keV.  The observed event rate as a function of $E'$ is then,
\begin{equation}\label{eq:diffrate_ep}
\frac{d R}{d E'} = \sum_T \int_0^\infty d E_R \, G_T(E',E_R) \, \epsilon(E_R) \, \frac{d R_T}{d E_R} \, ,
\end{equation}
where ${d R_T}/{d E_R}$ is the scattering rate per nuclide $T$ in eq.~\eqref{Eq: dRdER}.

For $\epsilon(E_R)$ we use the curve given by LZ in figure~S2\footnote{In this figure, the efficiency is given in terms of true nuclear recoil energy, as explained in ref.~\cite{aalbers2023search}.} of the Supplemental Material of ref.~\cite{LZ:2026axp}. For $G_T$ we use a Gaussian function
\begin{equation}
G_T(E', E_R) = \frac{1}{\sqrt{2\pi}\sigma(E_R)} \exp\left( -\frac{(E' - E_R)^2}{2\sigma(E_R)^2} \right)~,
\end{equation}
where $\sigma(E_R)$ is the energy resolution, parametrized as $\sigma/E_R=\sqrt{a^2/E_R+b^2+c^2}$. The first two terms describe the fluctuations modeled by NEST~\cite{Szydagis:2011tk}. We obtain them by simulating mono-energetic nuclear recoils at the position of the event, using the LZ detector model~\cite{LZ:2024zvo} with the gains $g_1$, $g_2$, the NR yield model of ref.~\cite{LZ:2026axp}, and the NR fluctuation parameters released with ref.~\cite{LZ:2026SolarNeutrinos}. Fitting the widths of the simulated peaks in the $E_R=[5.4, 270]$~keV region of interest (ROI)  where $\epsilon(E_R)>$ 50\%,  gives $a=0.575$~keV$^{1/2}$ and $b=0.0832$. The constant $c=0.1$ accounts for systematic effects not included in the simulation, and is chosen to approximately match the 23~keV systematic uncertainty on the event energy quoted by LZ~\cite{LZ:2026axp}. For the event we take $E_R=248$~keV, and the energy resolution is  $\sigma$(248 keV)$\simeq33$~keV.

We use an extended likelihood function of the form
\begin{equation}
\mathcal{L} = \frac{e^{-(N_\chi + \beta B)}}{N_{\rm obs}!} \times \prod_{j=1}^{N_{\rm obs}}\left[ MT\,\frac{d R}{d  E'}\biggr|_{E'=E'_j} + \frac{\beta B}{\Delta E'}\right] \times \frac{e^{-(\beta-1)^2/2\sigma_\beta^2}}{\sqrt{2 \pi}\sigma_\beta} \,,
\end{equation}
where $E'_j$ is the reconstructed energy of the $j$-th observed event, $N_{\rm obs}$ is the total number of observed events, $MT$ is the exposure, $d R/d E'$ is the DM differential scattering rate 
of eq.~\eqref{eq:diffrate_ep}. $N_\chi$ is the 
expected number of DM events and $\beta B$ is the expected number of background events, taken to be uniformly distributed over its width $\Delta E'$ defined by the numerical integration range from 3.9 to 306 keV, which corresponds to the ROI in $E_R$
extended by $1\sigma$ computed at the ROI boundaries on both sides.
The parameter $\beta$ is the background normalization, our only nuisance parameter ($\nu=\beta$), with a Gaussian constraint of relative width $\sigma_\beta$. We take the reference background  $B=0.01$ and $\sigma_\beta=7.5\%$, from the total background of $0.0106\pm0.0008$ counts estimated by LZ in the high-energy region S1c~$>500$~phd that contains the event~\cite{LZ:2026axp}. For the LZ data, $N_{\rm obs}=1$, with the single event at $E'=248$~keV.

In our statistical analysis we use the 2-sided profile likelihood ratio statistic 
\begin{equation}
t_\mu \equiv - 2 \ln \left( \frac{\mathcal{L}(\mu, \hat{\hat{\nu}})}{\mathcal{L}(\hat{\mu}, \hat{\nu})} \right) \, .
\end{equation}
Here, $\mu$ is our parameter of interest, a constant the DM scattering cross section is proportional to and characterizes its size, $\nu =\beta$ is the nuisance parameter, the double-hat refers to the parameter values that maximize the likelihood subject to the imposed constraint (namely a particular $\mu$ value), while the single hat denotes the unconditional maximization, with $\mu$ restricted to the physical region $\mu \geq 0$.
This is used to obtain the preferred parameter regions. Since the asymptotic $\chi^2$ distribution does not apply to a single observed event, we obtain the distribution of $t_\mu$ at each tested $\mu$ from toy Monte Carlo pseudo-experiments generated with $\beta=\hat{\hat{\beta}}$ and a fluctuated constraint on $\beta$, and include $\mu$ in the region at a given confidence level (CL) if the fraction of toys with $t_\mu$ at least as large as the observed value exceeds $1-{\rm CL}$~\cite{Cowan:2010js,Baxter:2021pqo}.

We use $t_\mu$ to produce preferred regions in DM parameter space. To construct the tables of local significand of the LZ event, we use instead the background-only statistic usually called $q_0$  defined as $q_0 = t_0$ if $\hat{\mu} \geq 0$, and 
$q_0 = 0$ if $\hat{\mu} < 0$, where $t_0$ is $t_\mu$ for $\mu=0$. Its distribution is likewise obtained from background-only toy Monte Carlo pseudo-experiments, and the local significance is $Z=\Phi^{-1}(1-p)$, where $p$ is the fraction of toys with $q_0$ at least as large as the observed value and $\Phi$ is the standard normal cumulative distribution.

\section{Significance of the LZ event across NREFT operators}
\label{sec:region}

The LZ collaboration assessed the significance of the 248~keV event for the elastic interactions for different NREFT operators and for inelastic endothermic scattering through the operators $\mathcal{O}_1$ and $\mathcal{O}_4$, in both cases assuming the SHM~\cite{LZ:2026axp}. Since the preferred parameter space for inelastic endothermic scattering lies close to the kinematic endpoint of the velocity distribution, both the choice of operator and the shape of the high-speed tail can significantly affect the inferred significance. Here we extend the LZ analysis in two directions. We consider inelastic endothermic scattering through each of the NREFT operators discussed in section~\ref{sec:eft}, for both isoscalar and isovector couplings, and we compute the significance of the event for different  DM masses and mass splittings $\delta>0$ for both the SHM and the MW+LMC velocity distributions introduced in section~\ref{sec:lmc}. We compute the recoil event rates using the publicly available code
\texttt{DDCalc}~\cite{GAMBITDarkMatterWorkgroup:2017fax}. For both the SHM and MW+LMC, we take a local DM density of $\rho_\chi = 0.3$~GeV\,cm$^{-3}$, which is the value recommended in ref.~\cite{Baxter:2021pqo} and used in the LZ analysis~\cite{LZ:2026axp}. 

\begin{figure}[t]
    \centering
    \includegraphics[width=\textwidth]{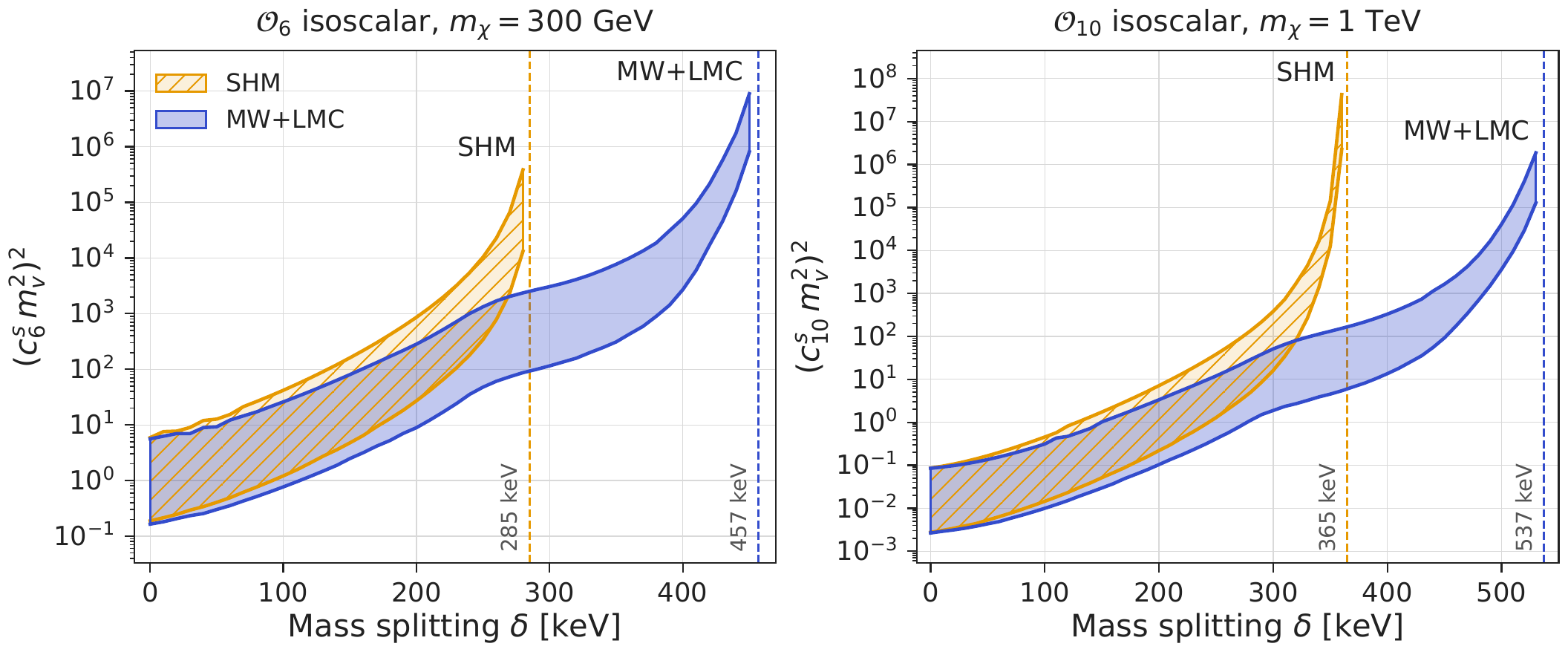}
    \caption{Effect of the LMC on the inelastic interpretation
    of the LZ 248~keV event for isoscalar interactions with $\mathcal{O}_6$ for
    $m_\chi=300$~GeV (left) and $\mathcal{O}_{10}$ for $m_\chi = 1$~TeV (right). The bands show the 90\% confidence
  intervals on the dimensionless coupling $(c_i^s\, m_v^2)^2$ as a
  function of the mass splitting, $\delta$, for the SHM (orange hatched)
  and the simulated MW+LMC  (blue shaded), where
  $m_v = 246.2$~GeV is the Higgs vacuum expectation value.  
    Dashed vertical lines mark the kinematic limits beyond which a recoil of energy $E_R=$ 281~keV, 1$\sigma$ above the LZ event reconstructed energy,  
    cannot be produced: $\delta = 285$~keV (SHM) and
  457~keV (MW+LMC) in the left panel, and $365$~keV (SHM) and $537$~keV
  (MW+LMC) in the right panel.
}
    \label{fig:NREFT_region}
\end{figure}

To illustrate how the LMC affects the preferred DM parameter space, figure~\ref{fig:NREFT_region} shows the 90\% CL intervals of the isoscalar coupling as a function of $\delta$ for two representative cases: $\mathcal{O}_6$ operator with $m_\chi = 300$~GeV (left panel) panel and $\mathcal{O}_{10}$ operator with $m_\chi = 1$~TeV (right panel). The interaction conventions and statistical
construction are described in sections~\ref{sec:eft} and~\ref{sec:stat},
respectively.
For both operators, the LMC shifts the allowed region to larger mass splittings, extending the range of splittings with an allowed signal interpretation. This follows directly from the kinematics. For a fixed recoil energy, the minimum speed required for endothermic scattering increases with $\delta$, and the largest mass splitting that can still produce the observed event is set by the maximum DM speed in the detector frame. The high-speed particles associated with the LMC extend the velocity distribution beyond the SHM endpoint, raising this kinematic limit from $\delta \simeq 285$~keV to $\delta \simeq 457$~keV for  $m_\chi = 300$~GeV and from $\delta \simeq 365$~keV to $\simeq 537$~keV for $m_\chi = 1$~TeV. At a fixed splitting below the SHM kinematic cutoff, the MW+LMC halo also requires a smaller coupling to reproduce the event, since more particles lie above $v_{\rm min}$.

\begin{figure}[t]
    \centering
        \includegraphics[width=0.495\textwidth]{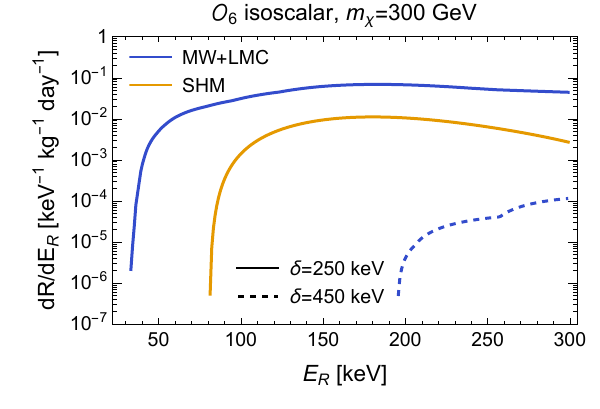}
        \includegraphics[width=0.495
        \textwidth]{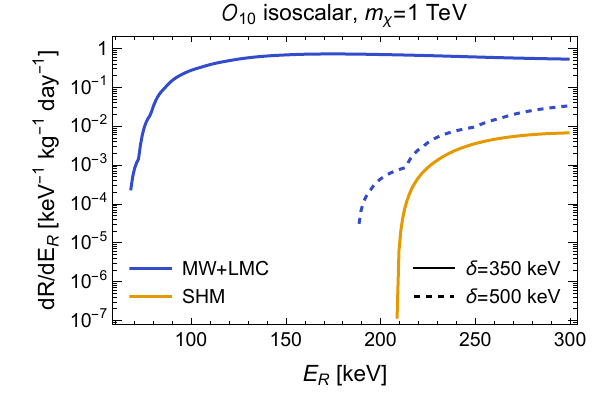} 
    \caption{Differential event rate as a function of recoil energy, $E_R$, for isoscalar interactions with  $\mathcal{O}_6$ and $m_\chi=300$~GeV (left) and  $\mathcal{O}_{10}$ and $m_\chi=1$~TeV (right), for the SHM (orange) and the MW+LMC (blue). In the left panel, $c_6^s=1$, with  solid curves  corresponding to $\delta=250$~keV and the dashed curve corresponding to $\delta=450$~keV. In the right panel, $c_{10}^s=1$, with solid curves for $\delta=350$~keV and the dashed curve for $\delta=500$~keV.
    For the higher value of $\delta$ in each panel, the event rate is zero for the SHM.}
\label{fig:rates_NREFT}
\end{figure}

Figure~\ref{fig:rates_NREFT} shows the corresponding differential event rates for the two benchmark cases and two $\delta$ values, with the isoscalar couplings set to $c_i^s=1$. Unlike elastic scattering, where the rate falls monotonically with recoil energy, inelastic endothermic scattering suppresses the spectrum at low recoil energies. For the smaller splitting (solid curves), both the SHM and MW+LMC produce events, but near the peak of the spectrum the MW+LMC halo enhances the rate by almost an order of magnitude for $\mathcal{O}_6$ with $m_\chi=300$~GeV, and by almost two orders of magnitude for $\mathcal{O}_{10}$ with $m_\chi=1$~TeV. The LMC also changes the shape of the spectrum. Since the minimum speed needed for a recoil is large at both low and high $E_R$, only a finite range of recoil energies is accessible, and this range widens as the maximum DM speed increases. The faster particles of the MW+LMC therefore extend the spectrum in both directions, most visibly to lower recoil energies. The difference between the SHM and MW+LMC rates grows with $\delta$. For the larger splittings (dashed curves), $v_{\rm min}$ exceeds the maximum DM speed in the SHM at every recoil energy, so the SHM rate vanishes, while the MW+LMC halo still predicts a non-zero rate. The high-$\delta$ region of the allowed parameter space in figure~\ref{fig:rates_NREFT} is therefore accessible only when the LMC is included.

Tables~\ref{tab:significance_m200}--\ref{tab:significance_m4000} show the
local significance of the LZ event for $m_\chi = 0.2$, 0.3, 0.4, 1.0 and 4.0~TeV for each operator and mass splitting with isoscalar (left panels) and isovector (right panels) interactions, for the SHM (top panels) and the simulated MW+LMC (bottom panels). Hatched cells indicate that no signal is expected in the analysis window, while cells marked zero correspond to a best-fit signal of zero significance.

For elastic scattering ($\delta = 0$), the significance is largely unaffected by the LMC, since a 248~keV recoil can be produced by DM particles much below the Galactic escape speed, where the SHM and the MW+LMC halo are similar. For inelastic scattering, the LMC extends the range of mass splittings for
which a signal is expected, and shifts the largest significance to larger
$\delta$. At $m_\chi = 1$~TeV, for example, the maximum significance occurs
at $\delta \sim 350$~keV in the SHM, but at $\delta \sim 500$~keV in the
MW+LMC analogue. The LMC does not, however, increase the significance for every interaction and splitting, since the enhanced high-speed tail also raises the predicted rate at other recoil energies, where no events were
observed.

The effect of the LMC on the maximum significance depends on the DM mass.
For $m_\chi = 200$--400~GeV, the largest significance in the MW+LMC halo
exceeds the largest in the SHM, most markedly at 200 and 300~GeV, where it
reaches $Z = 3.2$ and 3.3 compared with 2.6 and 2.8 in the SHM. Lighter DM requires higher speeds to produce a 248~keV recoil, so its signal depends more strongly on the high-speed tail where the LMC contribution is largest. At 1 and 4~TeV, the maximum significances in the SHM and MW+LMC agree within the Monte Carlo uncertainty of about 0.04, with $Z = 3.3$ at both these masses. For these masses the LMC changes where the event is best explained, moving this point to larger splittings, but not how well it is explained.

\begin{table}[p]
    \centering
    \includegraphics[width=\textwidth,height=0.8\textheight,keepaspectratio]
        {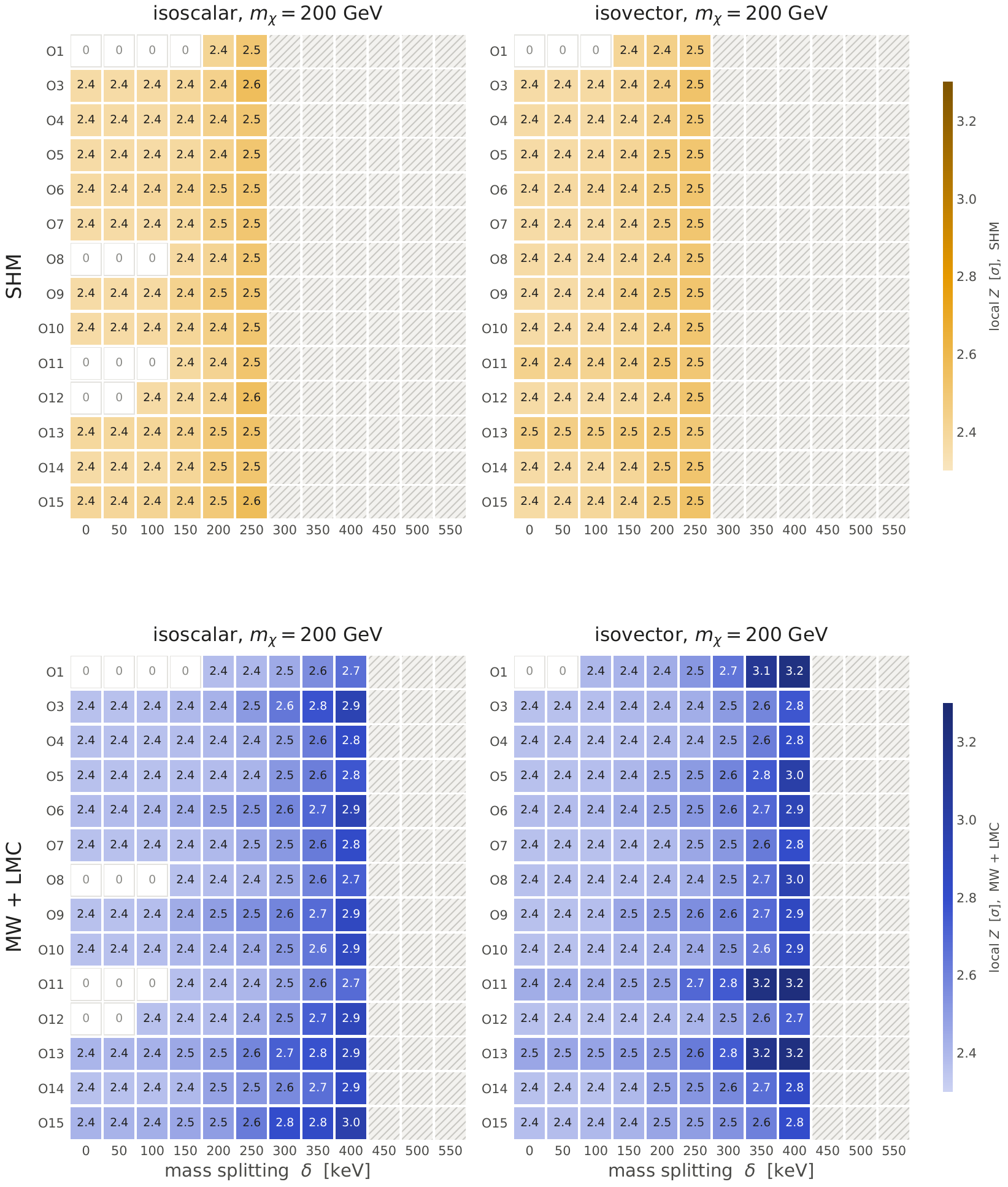}
    \caption{Local significance $Z$ of the LZ 248~keV event for
  $m_\chi = 200$~GeV, for each operator and mass splitting $\delta$, with
  the signal normalization as the only free parameter. The left and right
  panels show isoscalar and isovector interactions, and the upper (orange)
  and lower (blue) panels show the SHM and the simulated MW+LMC halo, on a
  common colour scale. Hatched cells have no expected signal in the ROI, and
  cells marked zero have a best-fit signal of zero ($q_0 = 0$). Values are
  given to two significant figures. The largest values are $Z = 2.6$ in the
  SHM, for isoscalar operators at $\delta=250$~keV, and $Z = 3.2$ in the MW+LMC for isovector operators at $\delta=350$--400~keV.
    }
    \label{tab:significance_m200}
\end{table}

\begin{table}[p]
    \centering
    \includegraphics[width=\textwidth,height=0.8\textheight,keepaspectratio]
        {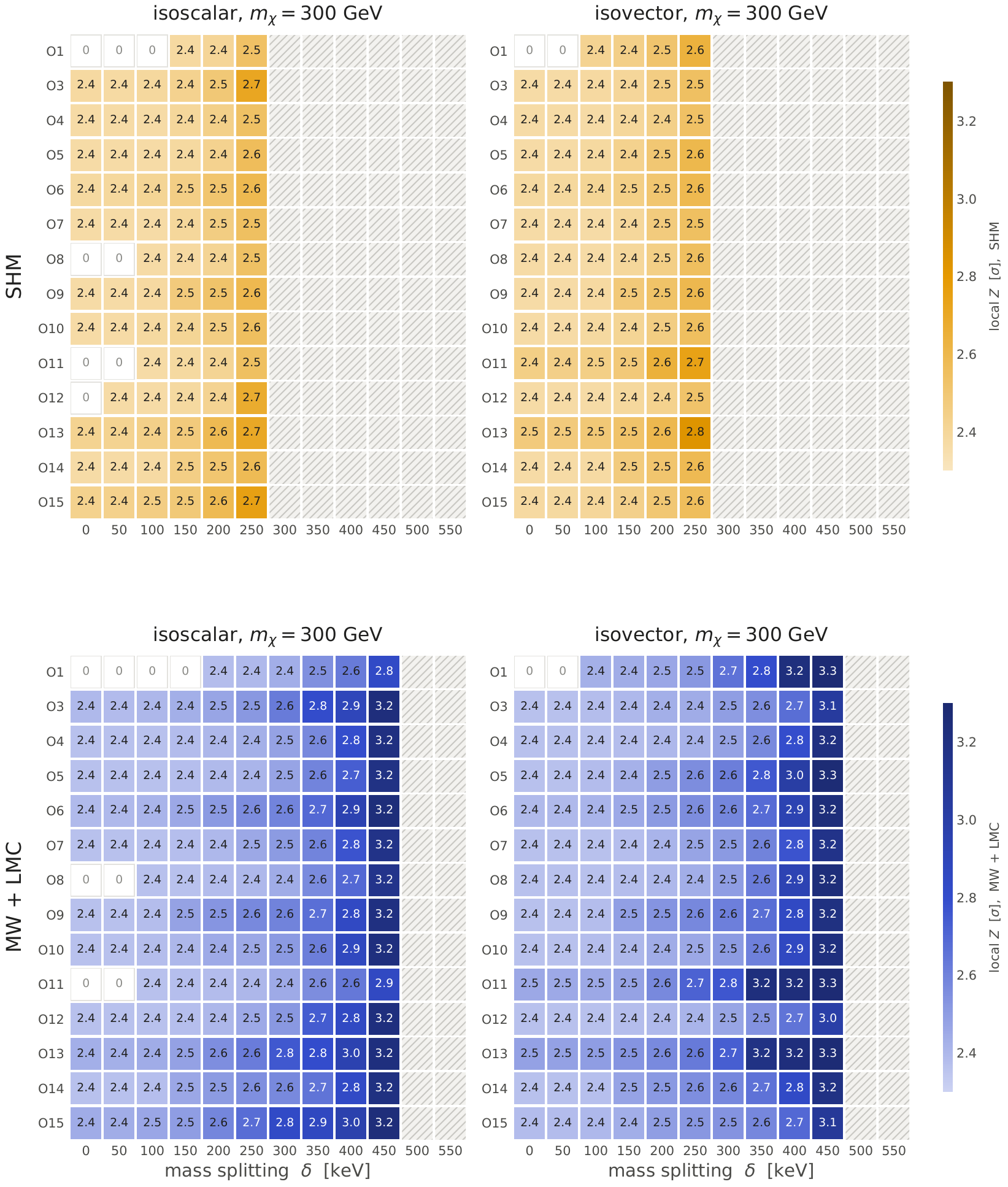}
    \caption{Same as table~\ref{tab:significance_m200}, for
    $m_\chi=300$~GeV. The largest values are $Z=2.8$ in the SHM, for
    isovector $\mathcal{O}_{13}$ at $\delta=250$~keV, and $Z=3.3$ in the
    MW+LMC halo, for isovector operators at $\delta=450$~keV.}
    \label{tab:significance_m300}
\end{table}

\begin{table}[p]
    \centering
    \includegraphics[width=\textwidth,height=0.8\textheight,keepaspectratio]
        {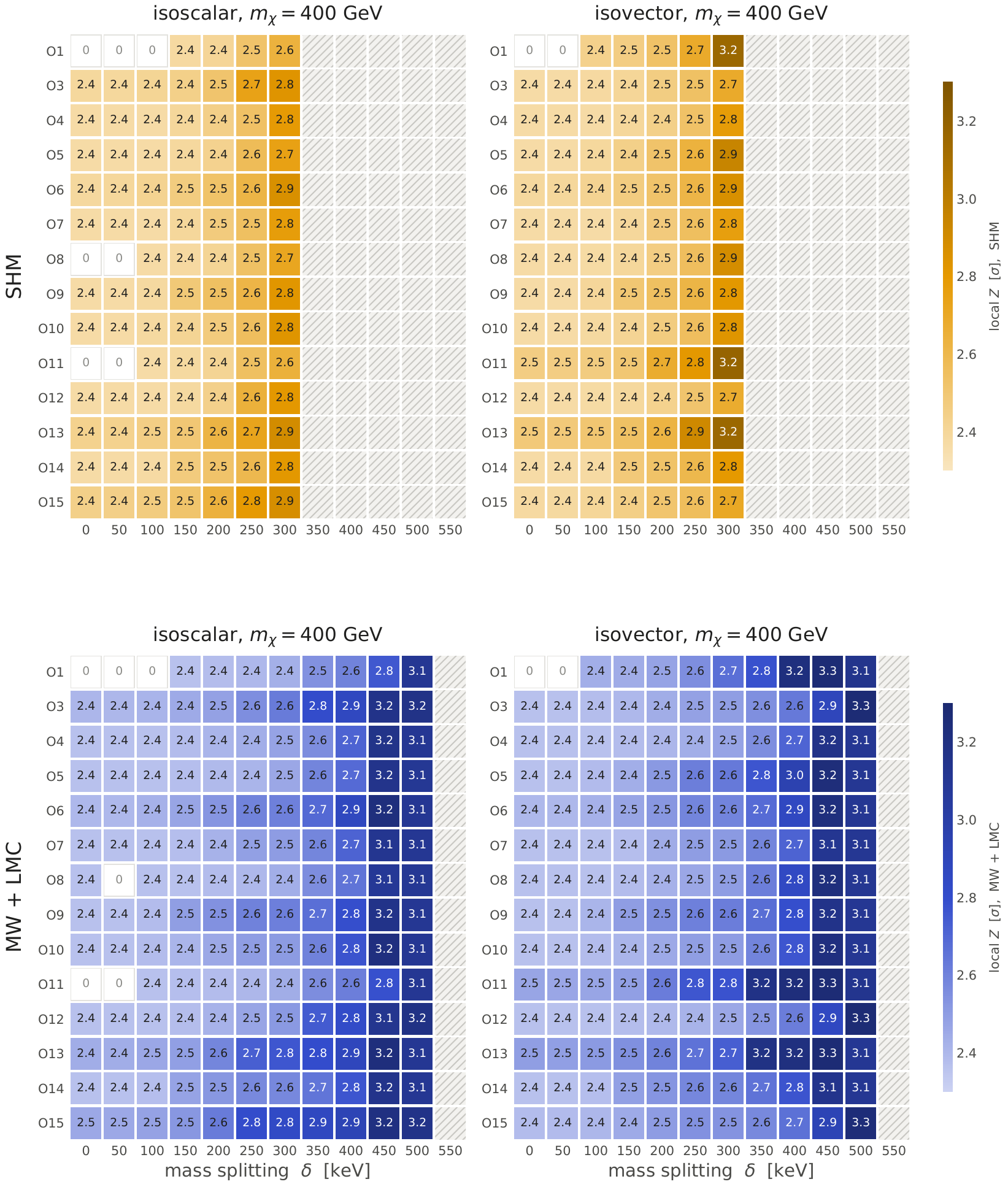}
    \caption{Same as table~\ref{tab:significance_m200}, for
    $m_\chi=400$~GeV. The largest values are $Z=3.2$ in the SHM, for
    isovector operators at $\delta=300$~keV, and $Z=3.3$ in the MW+LMC
    halo, for isovector operators at $\delta=450$--500~keV.}
    \label{tab:significance_m400}
\end{table}

\begin{table}[htbp]
    \centering
 \includegraphics[width=\textwidth,height=0.8\textheight,keepaspectratio]
        {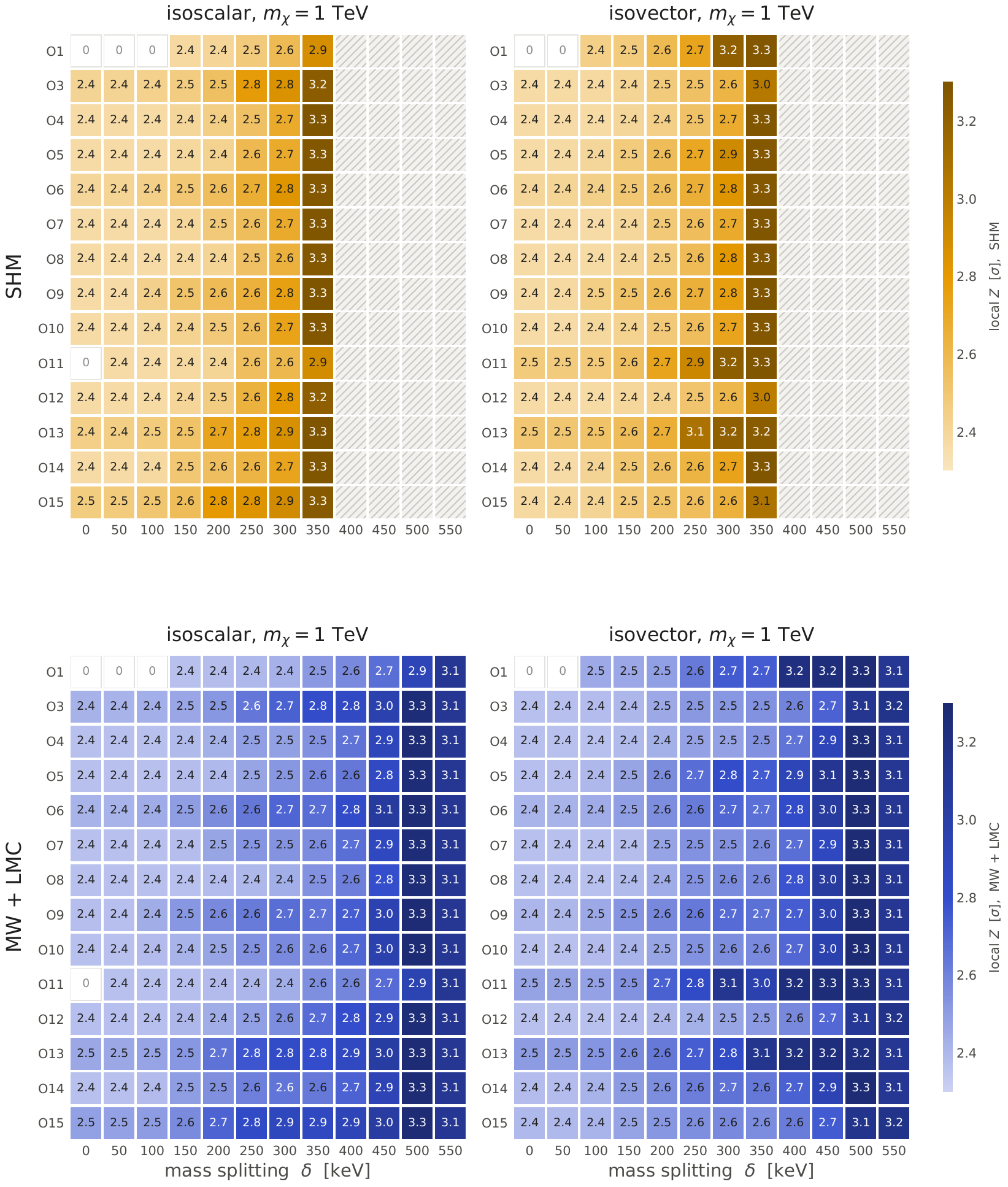}
    \caption{Same as table~\ref{tab:significance_m200}, for
    $m_\chi=1$~TeV. The largest values, $Z=3.3$, occur at $\delta=350$~keV in the SHM and at $\delta=450$--500~keV in the MW+LMC halo. The tables ends at $\delta = 550$~keV, since for larger splittings no signal is kinematically allowed for the MW+LMC halo either.
    }
    \label{fig:significance_m1000}
    \end{table}

\begin{table}[p]
    \centering
    \includegraphics[width=\textwidth,height=0.8\textheight,keepaspectratio]
        {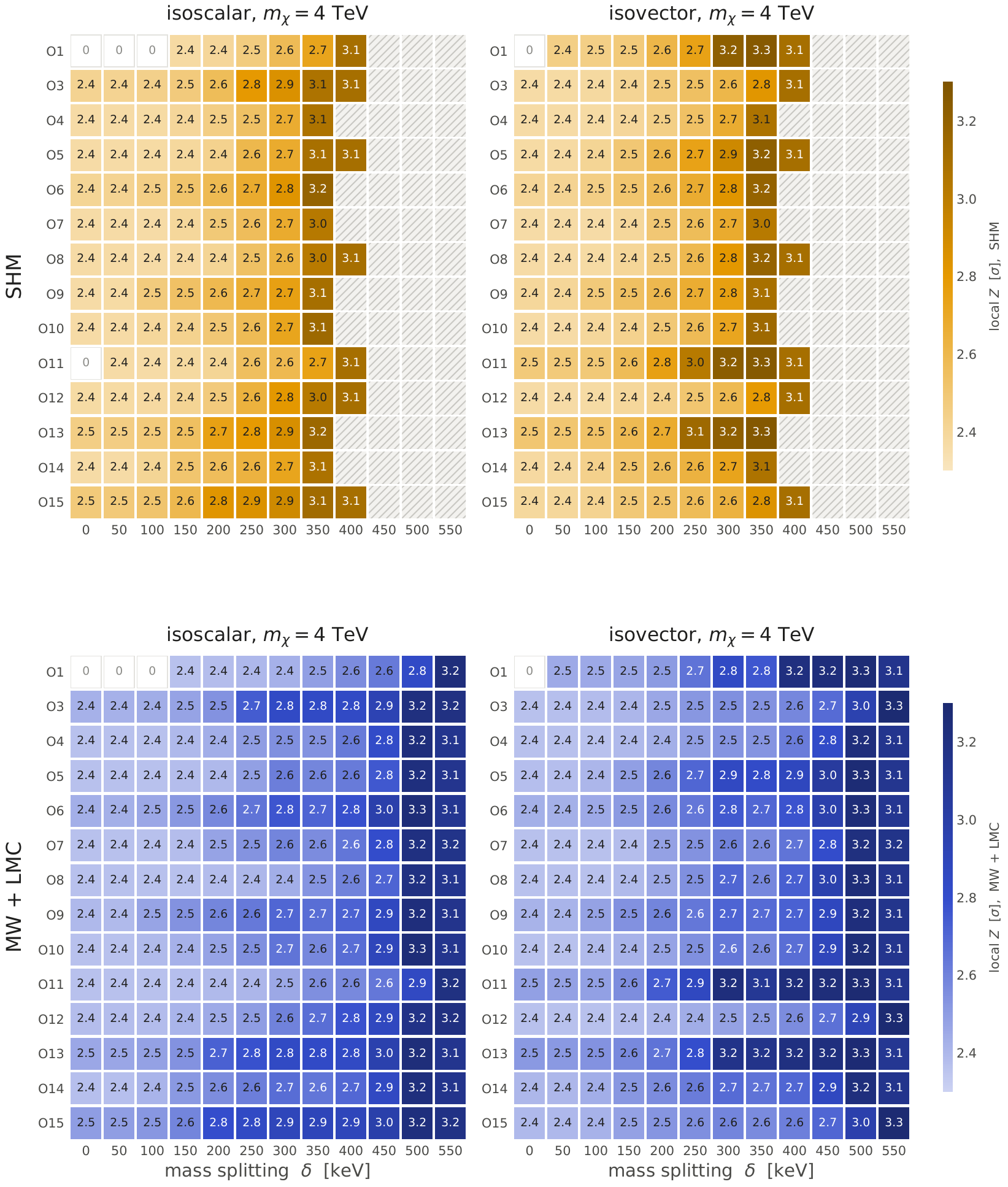}
    \caption{Same as table~\ref{tab:significance_m200}, for
    $m_\chi=4$~TeV. The largest values are $Z=3.3$ in the SHM, for
    isovector operators at $\delta=350$~keV, and $Z=3.3$ in the MW+LMC
    halo, at $\delta=500$--550~keV.
 The tables ends at $\delta = 550$~keV, since for larger splittings no signal is kinematically allowed for the MW+LMC halo either.}
    \label{tab:significance_m4000}
\end{table}

\afterpage{\clearpage}

\section{Singlet-Doublet dark matter}
\label{sec:SDM}

This section is motivated by the viable fit to the LZ event that a simple elastic SD collision provides, as shown by the 2.4 significance of the $\mathcal{O}_4$ operators in tables~\ref{tab:significance_m200}--\ref{tab:significance_m4000} (for $\delta=0$). As the preferred regions for inelastic endothermic scattering move to higher mass splitting values with the realistic MW+LMC halo, making them more susceptible to collider and indirect detection limits, a simple UV complete particle model which can provide the right DM abundance and be allowed by collider and inelastic limits becomes more appealing.

The Singlet-Doublet model  is a simple extension of the Standard Model (SM) (see e.g., \cite{Cohen:2011ec, Kearney:2016rng,Bhattiprolu:2025beq} and references therein). This model adds three colorless Weyl fermions: one electroweak singlet $S$, and two electroweak doublets $D$ and $D'$ with opposite weak-hypercharges to keep gauge anomalies canceled. The electric charges of the doublet components are $(0, -1)$ for $D$ and $(+1, 0)$ for $D'$. All three new fields are odd under a $Z_2$ symmetry, while all SM fields are even. Because of this parity,  doublets and singlet can couple with the Higgs only in  terms  $SHD$, and $SHD'$ with Yukawa coupling constants $\lambda$ and $\lambda'$, respectively. The singlet has a Majorana mass term  with mass $M_S$, while both doublets combine into a Dirac mass term with mass $M_D$. Therefore, the Singlet-Doublet model has four parameters beyond those of the SM, which for simplicity are taken to be real 
\begin{equation}
M_S, M_D, \lambda, \lambda'~.
\end{equation}
In this economical model the DM candidate  is the lightest neutral Majorana  mass eigenstate, consisting of an admixture of the singlet and doublet components.  It is similar to a mixed bino-higgsino DM state in the MSSM, or a singlino-higgsino in the NMSSM,  but without the mass and coupling relations imposed by supersymmetry.  

In these models the DM can be produced thermally~\cite{Cohen:2011ec, Kearney:2016rng,Bhattiprolu:2025beq,Paul:2025spm}. Requiring the correct relic abundance and applying 2025 direct detection constraints, Ref.~\cite{Bhattiprolu:2025beq} 
found that for $M_S < 850$~GeV, the DM state is predominantly composed of the singlet and has a mass  $m_\chi\simeq M_S < M_D$. For higher masses, the DM mass is $\simeq M_D$ and the DM state consists mostly of doublet components, approaching a higgsino, which was recently proposed as a candidate to explain the LZ event via inelastic scattering (see e.g. Refs.~\cite{Freese:2026sga,Wang:2026ytg}). 

We concentrate on the lighter DM candidate scattering elastically, and chosen to be at the \emph{Higgs blind spot}(Hbs), where the  coupling of the Majorana DM candidate to the Higgs boson vanishes. For this lighter candidate, the compressed mass spectrum of the four new Singlet-Doublet model particles (shown schematically in figure~5 of ref.~\cite{Bhattiprolu:2025beq})  makes their detection at the LHC difficult, ensuring they remain allowed by current LHC limits~\cite{Bhattiprolu:2025beq,Paul:2025spm}. 

The Higgs blind spot requires 
$\lambda'_{\rm Hbs} = - \lambda ({M_S}/{M_D}) [1 \pm \sqrt{1 -({M_S}/{M_D})^2}]^{-1}$.
Because of the Majorana nature of the fermion the vector coupling to the $Z$ boson vanishes,  leaving only an axial coupling,  $\mathcal{L}_{Z\chi\chi}= - (g_2/2 \cos \theta_W) g_{A} Z_{\mu} \bar{\chi} \gamma^{\mu} \gamma_5 \chi$, where $g_2$ is the SU$_{\rm L}$(2) SM gauge coupling constant, $\theta_W$ is the electroweak mixing angle, and
\begin{equation}
g_A=\frac{1}{2}~ \frac{v^2 (M_D^2-m_\chi^2) ({\lambda'_{\rm Hbs}}^2-\lambda^2)}{(M_D^2-m_\chi^2)+v^2 \left[({\lambda'_{\rm Hbs}}^2 +\lambda^2)(M_D^2+m_\chi^2)+4\lambda \lambda'_{\rm Hbs} M_D m_\chi\right]}.
\end{equation}
Here $v= 246$ GeV is the Higgs VEV defined via  $\left<{H}\right> = {v}/{\sqrt{2}}$,  and $m_\chi$ is the DM mass determined by the mixing of $S$ and the neutral states in $D$ and $D'$ in the mass matrix.

At tree level,  the  scattering cross section of $\chi$ off a nucleon $N= p, n$ is therefore purely  SD. The corresponding effective nucleon coupling is
$a_N = g_A \Sigma_{q} I^{(q)}_3 \Delta q^{(N)}$, where $q=u,d,s$, and $I^{(q)}_3$ is the weak isospin component ($\pm 1/2$) of each quark flavor.  This yields $a_p = 0.675 g_A$ and $a_n = - 0.595 g_A$. The SD cross section is  $\sigma_{SD}^{\chi N} = (24/ \pi) G_F^2 \mu_{\chi N}^2 a_N^2$, where $\mu_{\chi N}$ is the DM-nucleon reduced mass. Therefore, using as reference the reduce masses of $m_\chi=100$~GeV, the SD cross sections of $\chi-$neutron and $\chi-$proton are $\sigma_{SD}^{\chi n}=  1.24 g_A^2 \times 10^{-37} (\mu_{\chi n}/ 0.9308 {\rm GeV})^2$~cm$^2$  and 
$\sigma_{SD}^{\chi p}=  1.59 g_A^2 \times 10^{-37} (\mu_{\chi p}/ 0.9295 {\rm GeV})^2$~cm$^2$.

\begin{figure}[t] 
    \centering   
     \includegraphics[width=\textwidth]{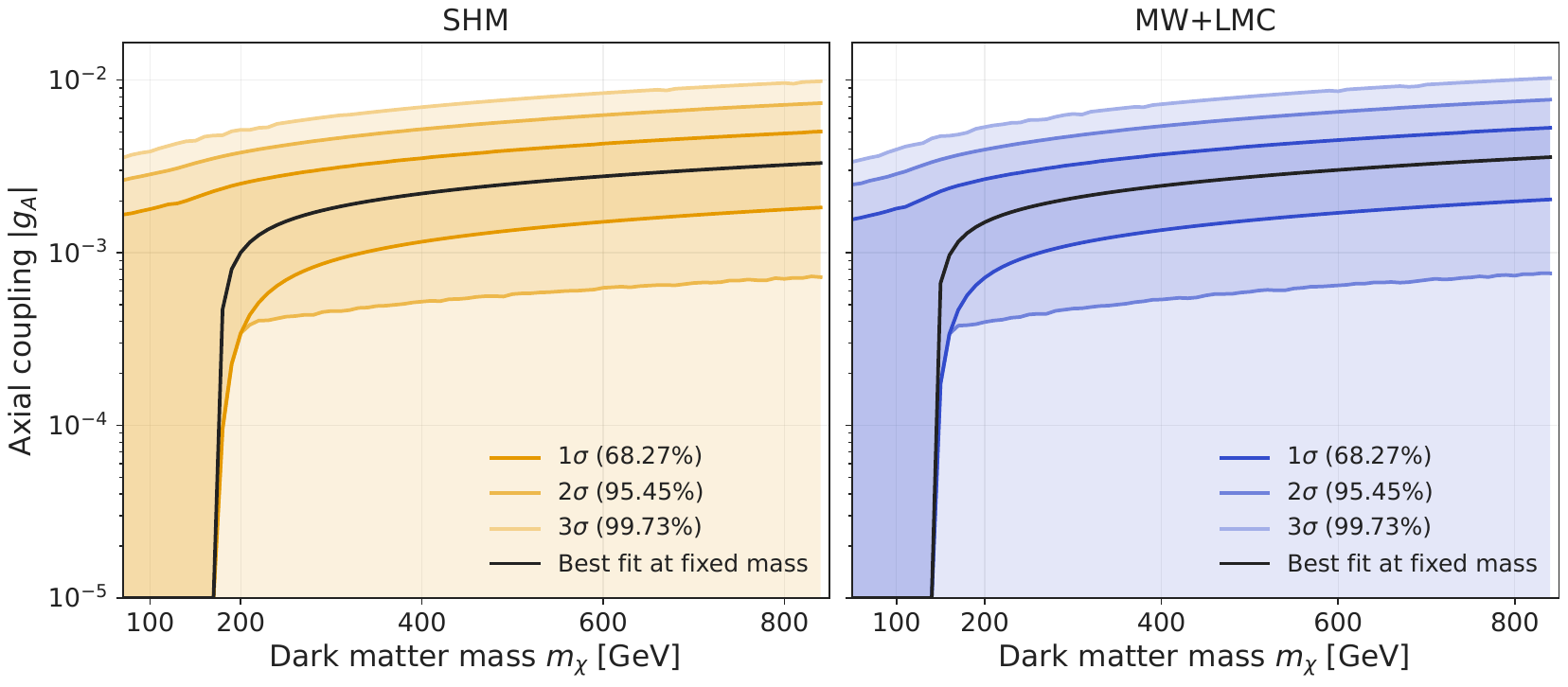}
    \caption{Axial coupling $|g_A|$  of the light Majorana Singlet-Doublet DM candidate to the $Z$ boson at the Higgs blind spot  as a function of the DM mass $m_\chi$. The black line and colored regions show the best fit and 1$\sigma$, 2$\sigma$ and 3$\sigma$ CL fits to the LZ event under the SHM (left) and the MW+LMC (right) halo model. The regions are cutoff at the minimum $m_\chi$ that can produce a recoil of energy $E_R=$ 215~keV, 1$\sigma$ below the LZ event reconstructed energy: $m_\chi \sim 70$~GeV for the SHM and $m_\chi \sim 50$~GeV for the MW+LMC. The favored mass range is $m_\chi\gtrsim 200$~GeV.}
    \label{fig:gA-SDM}
\end{figure}

In  figure~\ref{fig:gA-SDM}, we show the $|g_A|$ values as a function of the DM mass for which our Singlet-Doublet model candidate provides a good fit to the LZ event at the 1$\sigma$, 2$\sigma$ and 3$\sigma$ CL. The black line in this figure indicates the best fit as a function of mass. 
The favored mass range is $m_\chi> 200$~GeV and extends to the highest masses, $\simeq 850$ GeV, allowed by correct relic abundance and 2025 direct detection constraints~\cite{Bhattiprolu:2025beq}. 

\begin{figure}[t]
    \centering
        \includegraphics[width=0.6\textwidth]{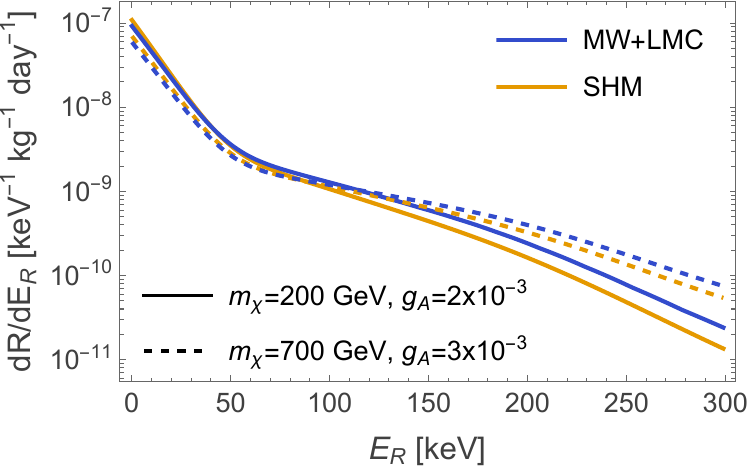} 
    \caption{Differential event rate as a function of recoil energy, $E_R$, for elastic SD scattering with $a_p=0.675$, $a_n=-0.595$, for the SHM (orange) and the MW+LMC analogue (blue). The solid curves correspond to a DM mass of $m_\chi=200$~GeV and $g_A=2 \times 10^{-3}$, while the dashed curves correspond to $m_\chi=700$~GeV and  $g_A=3 \times 10^{-3}$, corresponding to the best fit values for the axial coupling to the $Z$ boson of the Majorana Singlet-Doublet DM model at the Higgs blind spot.}
\label{fig:rate-SD}
\end{figure}

Two examples of the recoil rate of this candidate, shown in figure~\ref{fig:rate-SD} for $m_\chi=200$ and 700~GeV  with coupling $g_A$ chosen close to the respective best-fit values in figure~\ref{fig:gA-SDM}, demonstrate why this DM particle is compatible with the LZ event. The spectra are relatively flat (in comparison to other rejected models, e.g.~with SI elastic scattering corresponding to the $\mathcal{O}_1$ operator. However, They show a peak at low recoil energy, which would be compatible with LZ having future events at lower energies. This is in clear contrast to what would be expected for models with inelastic endothermic scattering, in which new events should not appear at low energies.

We have not yet addressed indirect detection  limits for this candidate. Fermi-LAT  limits  do not restrict this model due to the highly suppressed late-Universe self-annihilation cross section necessary for successful thermal production within a compressed dark sector spectrum~\cite{Kearney:2016rng,Bhattiprolu:2025beq}. Such limits are only relevant if the DM candidate would have a much larger annihilation cross section, which would render it underabundant if thermally produced~\cite{Calibbi:2015nha}. Two identical Majorana fermions annihilate dominantly in a p-wave, where the cross section depends on their speed $v$,
$\sigma_{\rm ann} v \propto v^2$, and while $v\simeq$ 0.3 at freeze-out, it is about $ v\simeq 10^{-3}$ at late times, heavily suppressing a Fermi-LAT signal.

Severe indirect detection constraints arise instead from $\chi$ capture in the Sun, due to  $\chi-$proton SD scattering.  Subsequent annihilation in the solar core would produce 
high-energy neutrinos that IceCube could detect. The strongest IceCube limits comes from the $W^+W^-$ annihilation channel~\cite{IceCube:2025fcu}. In the Sun's core, DM particles would  have thermalized speeds $v\simeq 10^{-3}$; thus, only s-wave annihilation is important.  The direct s-wave annihilation of our Majorana fermions into a $W^+ W^-$ pair is forbidden. It is blocked for s-channel $Z$ exchange due to quantum number conservation, and for t-channel exchange of charged $\chi^{\pm}$ fermions because at the Higgs blind spot the necessary coupling to the longitudinal $W$ component (i.e.~a Goldstone boson component of the Higgs field) vanishes~\cite{Kearney:2016rng}. 

However, energetic $W$ bosons with a broad energy spectrum and average energy $ \left<E_W\right>\simeq 0.6~m_\chi$ can be produced via annihilation into top-quark pairs, followed by the rapid  decay $t\to bW$~\cite{Cohen:2011ec,Kearney:2012rf}. S-wave annihilation into fermions ($\chi \chi \to f \bar{f}$)  is  $\propto (m_f/m_Z)^2$,  and thus unsuppressed for top quarks. The IceCube upper limit~\cite{IceCube:2025fcu} assumes direct annihilation into  two monoenergetic  $W$ bosons with $E_W= m_\chi$, rather than  the broader energy spectrum in the $ t\bar{t} \to W^+b W^- \bar{b}$ mode. While this annihilation mode has been discussed~\cite{Cohen:2011ec,Kearney:2012rf}, its corresponding  IceCube limit has not been computed for our candidate in the  $m_\chi > m_{\rm top}$ regime. Based on calculations done for $m_\chi < m_{\rm top}$, ref.~\cite{Kearney:2012rf}  estimated that the IceCube scattering cross section limit for $m_\chi > m_{\rm top}$ softens by a factor of up to four. 

A rough estimate of the IceCube limit  softening factor can be done using simple scaling arguments. In the Sun, for the cross section considered, equilibrium has been achieved, and thus the annihilation rate is equal to half the capture rate. The capture rate  scales with the DM mass as $1/m_\chi^2$, because both the $\chi$ number density and the energy loss per proton collision are $\propto 1/m_\chi^2$. Since the number of neutrinos produced per annihilation through $W$ boson decays depends only on $E_W$, we can estimate the shift of the IceCube SD scattering cross section upper limit that applies to the $W^+b W^- \bar{b}$ mode. A rough estimate consists of shifting the $WW$ limit line to higher masses by a factor $1/0.6\simeq 1.7$, and upward to lower cross sections by a factor of $0.6^2$. For $m_\chi <800$~GeV,  this procedure softens the limit by approximately a factor of about 5. This would not exclude the lower portion of the 200~GeV$ \lesssim m_\chi \lesssim 800$~GeV mass range of the light Majorana Singlet-Doublet model DM candidate at the Higgs blid spot. However a proper calculation needs to be done to determine the actual viable mass range.

At the tree-level Higgs blind spot where we operate, one-loop corrections introduce a small coupling between the dark sector and the Higgs~\cite{Han:2018gej}. This induces a SI cross section roughly two orders of magnitude below the sensitivity of current direct detection experiments~\cite{Han:2018gej}. In the Sun, the loop corrections enable  direct s-wave  annihilation  into two longitudinal $W$ bosons, $\chi \chi \to W^+ W^-$,   whose cross section can be estimated to be roughly six orders of magnitude smaller than the tree-level $W^+b W^- \bar{b}$ annihilation  mode. This direct annihilation proceeds via t-channel exchange of  charged dark-sector fermions $\chi^{\pm}$, since the s-channel exchange of a Higgs boson only leads to p-wave annihilation. Therefore, the tree-level top-quark channel remains the dominant driver of solar signals for our DM candidate.

\section{Summary}
\label{sec:summary}

In this work, we have assessed how the  LMC affects the interpretation of the 248~keV nuclear recoil event reported by the LZ experiment, using a simulated MW+LMC analogue from the Auriga magneto-hydrodynamical simulations and comparing it with the SHM. The LZ event has motivated a large number of interpretations in terms of endothermic inelastic scattering, in which the mass splitting $\delta$ between the DM ground state and its excited state suppresses low-energy recoils. Since the splitting increases the minimum DM speed required to produce a recoil, these interpretations rely on the fastest DM particles in the Solar neighborhood, and are therefore sensitive to the high-speed tail of the local DM velocity distribution.

We computed the local significance of the LZ event for all NREFT operators, considering elastic and inelastic scattering and both isoscalar and isovector couplings, for DM masses of 0.2, 0.3, 0.4, 1 and 4~TeV. For elastic scattering, the significance is largely unaffected by the LMC. For inelastic scattering, the LMC extends the range of mass splittings that can produce a signal. Splittings for which the SHM predicts no events in the LZ analysis window become accessible in the MW+LMC halo. As a result, the best fit to the event moves to higher $\delta$, from 250~keV to 450~keV at $m_\chi = 300$~GeV and from 350~keV to 450--500~keV at $m_\chi = 1$~TeV. The impact of the LMC on the quality of the fit varies between interactions, but depends most strongly on the DM mass. The improvement is largest for light DM, which must have higher speeds to deposit 248~keV in a xenon nucleus. The highest significance increases from 2.6 to 3.2 at 200~GeV and from 2.8 to 3.3 at 300~GeV. At 1 and 4~TeV, the highest significance is essentially the same in both the SHM and MW+LMC halo, so the LMC shifts the preferred parameter space without improving the fit.

These trends are reflected in the allowed regions and recoil spectra. For isoscalar $\mathcal{O}_6$ scattering with $m_\chi = 300$~GeV, the largest splitting that can produce the observed recoil increases from about 285~keV in the SHM to 457~keV in the MW+LMC halo, and a similar shift occurs for $\mathcal{O}_{10}$ with $m_\chi = 1$~TeV. For splittings accessible in both the SHM and MW+LMC, the LMC enhances the rate near the peak of the spectrum by about one and two orders of magnitude for the two benchmarks, respectively, and extends the spectrum to lower recoil energies.

As a simple UV-complete alternative, we also considered a relatively light Majorana Singlet-Doublet DM candidate at the Higgs blind spot, which scatters elastically through its axial coupling $g_A$ to the $Z$ boson. This model is motivated by two of our results. First, elastic SD scattering through $\mathcal{O}_4$ already provides a reasonable fit to the LZ event, with a local significance of 2.4. Second, since the MW+LMC halo moves the preferred regions for inelastic endothermic scattering to larger mass splittings, inelastic interpretations may face stronger constraints from collider and indirect searches, which makes an elastic explanation more appealing. We determined the values of $|g_A|$ for which the model fits the LZ event at the 1, 2 and 3$\sigma$ levels as a function of the DM mass. The favored masses are $m_\chi > 200$~GeV, extending up to $m_\chi \simeq 850$~GeV, the largest mass compatible with the correct relic abundance and current direct detection constraints (though the higher end of this range is severely challenged by IceCube limits).

Our results show that the LMC substantially changes the inelastic endothermic interpretation of the LZ event,  while elastic interpretations are largely unaffected. Since inelastic endothermic interpretations probe DM particles near the Galactic escape speed, they should be based on realistic velocity distributions that account for the impact of the LMC. With the full LZ exposure, the recoil spectrum of any additional events, together with searches using heavier targets, will help distinguish between inelastic interpretations of the event and elastic ones such as the Singlet--Doublet model considered here, and test the preferred parameter space identified.

\acknowledgments N.B.~acknowledges the support of the Canada Research Chairs Program, the Natural Sciences and Engineering Research Council of Canada (NSERC), funding reference number RGPIN-2020-07138, and the NSERC Discovery Launch Supplement, DGECR-2020-00231. S.C.~was partially supported by the National Science Foundation Graduate Research Fellowship Program. G.B.G.~was partially supported by the US Department of Energy under Award Number DE-SC0009937. A.C.K., S.C., and  Y.X.~acknowledge the support of the Department of Energy funding reference number DE-SC0025629, of the Alfred P. Sloan Research Fellowship, grant Number: FG-2025-24419, and of the Cottrell Research Foundation, funding reference number CS-CSA-2025-058. J.R.C.~acknowledges support from INFN through the Senior Research Fellowship program (Grant No.~27076). Y.X.~has received funding from the European Union's Horizon Europe research and innovation program under the Marie Skłodowska-Curie grant agreement No.~101126636.

\clearpage

\typeout{}
\bibliographystyle{JHEP}
\bibliography{refs}
\end{document}